\documentclass{edp-conf}
\usepackage{graphicx}
\def\inu{{I_{\nu}}}
\def\bnu{{B_{\nu}}}
\def\mic{$\mu$m}

\begin{document}

\TitreGlobal{SF2A 2003}

\title{Inverse temperature dependence of the dust submillimeter spectral index}
\author{Dupac, X.}\address{ESA-ESTEC, Keplerlaan 1, 2201 AZ Noordwijk, The Netherlands}
\author{Bernard, J.-P., Boudet, N., Giard, M.}\address{CESR, 9 av. du Colonel Roche, BP 4346, 31028 Toulouse cedex 4, France}
\author{Lamarre, J.-M.}\address{LERMA, Obs. de Paris, 61 av. de l'Observatoire, 75014 Paris, France}
\author{M\'eny, C.}\address{CESR, 9 av. du Colonel Roche, BP 4346, 31028 Toulouse cedex 4, France}
\author{Pajot, F.}\address{IAS, Campus d'Orsay, b\^at. 121, 15 rue Clemenceau, 91405 Orsay cedex, France}
\author{Ristorcelli, I.}\address{CESR, 9 av. du Colonel Roche, BP 4346, 31028 Toulouse cedex 4, France}
\runningtitle{Temperature dependence of the dust spectral index}
\setcounter{page}{237}
\index{Author1, A.}
\index{Author2, B.}
\index{Author3, C.}

\maketitle
\begin{abstract}We present here a compilation of PRONAOS-based results concerning the temperature dependence of the dust submillimeter spectral index, including data from Galactic cirrus, star-forming regions and a circumstellar envelope.
We observe large variations of the spectral index (from 0.8 to 2.4) in a wide range of temperatures (12 to 80 K).
These spectral index variations follow a hyperbolic-shaped function of the temperature, large spectral indices (1.5-2.4) being observed in cold regions (12-20 K) while small indices (0.8-1.6) are observed in warm regions (35-80 K).
Three interpretations are proposed: one is that the grain sizes change in warm and rather dense environments, another is that the chemical composition of the grains is not the same in different environments, a third one is that there is an intrinsic dependence of the dust spectral index on the temperature due to quantum processes.
\end{abstract}
%
\section{Introduction}
To accurately characterize dust emissivity properties represents a major challenge of nowadays astronomy.
PRONAOS (PROgramme NAtional d'Observations Submillim\'etriques) is a French
balloon-borne submillimeter experiment (\cite{ristorcelli98}).
Its effective wavelengths are 200, 260, 360 and 580 \mic, and the angular resolutions are 2$'$ in bands 1 and 2, 2.5$'$ in band 3 and 3.5$'$ in band 4.

\section{Analysis of the Galactic dust emission}

We fit a modified black body law to the spectra: $\inu = \epsilon_0 \; \bnu(\lambda,T) \; (\lambda/\lambda_0)^{-\beta}$, where $\inu$ is the spectral intensity (MJy/sr), $\epsilon_0$ is the emissivity at $\lambda_0$ of the observed dust column density, $\bnu$ is the Planck function, $T$ is the temperature and $\beta$ is the spectral index.

\begin{figure}[!ht]
\includegraphics[scale=.4]{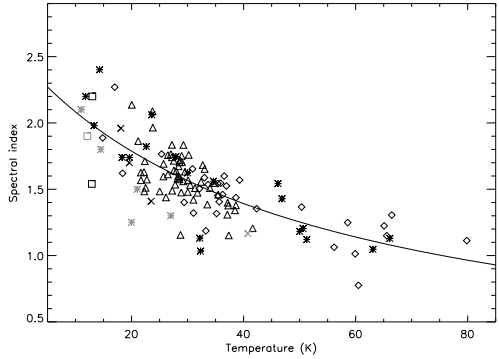}
\caption[]{Spectral index versus temperature, for fully independent pixels in Orion (black asterisks), M17 (diamonds), Cygnus (triangles), $\rho$ Ophiuchi (grey asterisks), Polaris (black squares), Taurus (grey square), NCS (grey cross) and NGC 891 (black crosses).
The full line is the result of the best hyperbolic fit: $\beta = {1 \over 0.4 + 0.008 T}$
}
\label{compil}
\end{figure}

We present in Fig. \ref{compil} the spectral index - temperature relation observed.
The temperature in this data set ranges from 11 to 80 K, and
the spectral index also exhibits large variations from 0.8 to 2.4.
One can observe an anticorrelation on these plots between the temperature and the spectral index, in the sense that the cold regions have high spectral indices around 2, and warmer regions have spectral indices below 1.5.
In particular, no data points with $T >$ 35 K and $\beta >$ 1.6 can be found, nor points with $T <$ 20 K and  $\beta <$ 1.5.
The temperature dependence of the emissivity spectral index is well fitted by a hyperbolic approximating function.
Several interpretations are possible for this effect: one is that the grain sizes change in dense environments, another is that the chemical composition of the grains is not the same in different environments and that this correlates to the temperature, a third one is that there is an intrinsic dependence of the spectral index on the temperature, due to quantum processes such as two-level tunneling effects.
Details about this analysis and the possible interpretations can be found in Dupac et al. (2003).

\end{document}